\documentclass[10pt, conference, compsocconf]{IEEEtran}

\usepackage{graphicx}

\begin{document}

\title{On Methods for the Formal Specification \\ of Fault Tolerant Systems}
\author{\IEEEauthorblockN{Manuel Mazzara}
\IEEEauthorblockA{School of Computing Science, Newcastle University, UK\\
Manuel.Mazzara@newcastle.ac.uk}
}

\maketitle

\begin{abstract} 
This paper introduces different views for understanding problems and faults with the goal of defining a method for the formal specification of systems. The idea of Layered Fault Tolerant Specification (LFTS) is proposed to make the method extensible to fault tolerant systems. The principle is layering the specification in different levels, the first one for the \textit{normal behavior} and the others for the \textit{abnormal}. The abnormal behavior is described in terms of an Error Injector (EI) which represents a model of the erroneous interference coming from the environment. This structure has been inspired by the notion of idealized fault tolerant component but the combination of LFTS and EI using Rely/Guarantee reasoning to describe their interaction can be considered as a novel contribution. The progress toward this method and this way to organize fault tolerant specifications has been made experimenting on case studies and an example is presented.
\end{abstract}


% A category with the (minimum) three required fields
%\category{D.2.1}{Software Engineering}{D.2.1 Requirements and Specifications}
%A category including the fourth, optional field follows...
%\category{F.4.3}{Mathematical Logic and Formal Languages}{Formal Languages}
%\terms{}

\section{Introduction}

%\begin{quote}
%\textit{``Dubium Sapientiae initium'' - Descartes}
%\end{quote}

There is a long tradition of approaching Requirements Engineering (RE) by means of formal or semi-formal techniques. Although "fuzzy" human skills are involved in the process of elicitation, analysis and specification - as in  any other human field - still methodology and formalisms can play an important role \cite{SMART}. However, the main RE problem has always been communication. A definition of communication teaches us that \cite{DeFleur93}:

\begin{quote}
``Human communication is a process during which source individuals initiate messages using conventionalized symbols, nonverbal signs, and contextual cues to express meanings by transmitting information in such a way that the receiving party constructs similar or parallel understanding or parties toward whom the messages are directed.''
\end{quote}

The first thing we have realized in building dependable software is that it is necessary to build dependable communication between parties that use different languages and vocabulary. In the above definition you can easily find the words \textit{"similar or parallel understanding are constructed by the receiving parties"}, but for building dependable systems matching expectations (and specification) it is not enough to build a \textit{similar or parallel understandings} since we want a more precise mapping between intentions and actions. Formal methods in system specification look to be an approachable solution.

Object Oriented Design \cite{BoochOO} and Component Computing \cite{SzyperskiCS} are just well known examples of how some rigor and discipline can improve the final quality of software artifacts besides the human communication factor. The success of languages like Java or C\# could be interpreted in this sense, as natural target languages for this way of structuring thinking and design. It is also true - and it is worth reminding it - that in many cases it has been the language and the available tools on the market that forced designers to adopt object orientation principles, for example, and not vice versa. This is the clear confirmation that it is always a combination of conceptual and software tools together that create the right environment for the success of a discipline.

Semi formal notations like UML \cite{FowlerUML} helped in creating a language that can be understood by both specialists and non specialists, providing different views of the system that can be negotiated between different stakeholders with different backgrounds. The power (and thus the limitation of UML) is the absence of a formal semantics (many attempts can be found in the literature anyway) and the strong commitment on a way of reasoning and structuring problems which is clearly the one disciplined by object orientation. Many other formal/mathematical notation existed for a long time for specifying and verifying systems like process algebras (a short history by Jos Baeten in \cite{BaetenPA}) or specification languages like Z (early description in \cite{AbrialZ-book}) and B \cite{AbrialB-book}. The Vienna Development Method (VDM) is maybe one of the first attempts to establish a Formal Method for the development of computer systems \cite{BjornerVDM78}. A survey on these (and others) formalisms can be found in \cite{MazzaraDEPEND2010}. All these notations are very specific and can be understood only by specialists. The point about all these formalisms is that they are indeed notations, formal or semi-formal. Behind each of them there is a way of structuring thinking that does not offer complete freedom and thus forces designers to adhere to some discipline. But still they are not methods in the proper sense, they are indeed languages. 

\subsection*{Contributions of the paper}

The goal of this paper is providing a different view for interpreting problems and faults. The overall result will be the definition of a method for the specification of systems that do not run in isolation but in the real, physical world. In \cite{mazzara:ewdc09} we mainly defined a draft of this approach contributing with an understanding of what a method is and an analysis of the desiderata. We then presented our method and its application to a Train System example. We realized that few points were still at a draft stage and their explanation still obscure in some paragraphs. In this paper we will provide more details instead and a different example. The main contributions of this work can be considered:

\begin{enumerate}
\item A perspective for describing problems in term of static view and dynamic view and and a discussion on how to combine them
\item A perspective to describe faults in terms of an Error Injector representing a model of faults (and consequent introduction of fault tolerant behavior)
\item The organization of the specification in terms of layers of Rely Guarantee conditions (LFTS)
\item The experimentation on a small automotive case study
\end{enumerate}

\section{An Angle to See Problems}

\begin{figure}[htp]
\centering
\includegraphics[scale=0.2]{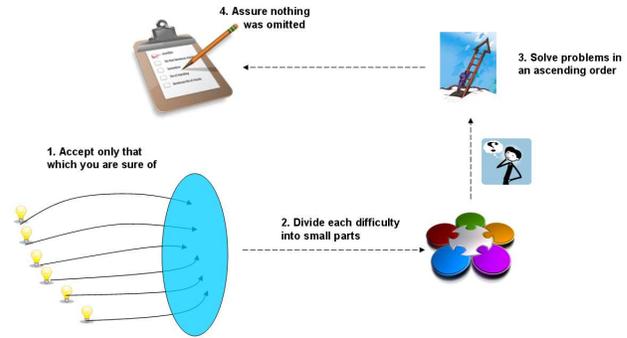}
\caption{The method of science}\label{descartes}
\end{figure}

Our work in this paper focuses especially on \cite{JonesHJ07} where the original idea of a formal method for the specification of systems running in the physical world originated. That paper was full of interesting ideas but still was lacking of a method in the sense we described in \cite{mazzara:ewdc09} and \cite{mazzara:tr09}. Few case studies have been analyzed according to this philosophy in \cite{Jones05t} but still a complete method has not been reached. For this reason we think now that a more structured approach is urgent in this area. Thus, the goal of the present work is improving our understanding of those ideas and incrementing that contribution putting it in an homogeneous and uniform way and describing a method featuring the properties we introduced in \cite{mazzara:ewdc09}, with particular attention to fault tolerance. In figure \ref{descartes} we report a graphical synthesis of the Descartes method presented in \cite{Descartes}. This work presents a method as consisting of a partially ordered set of actions which need to be performed and then discharged within a specific causal relationship. The success of one action determines the following ones. Furthermore, the method has to be repeatable, possibly by non experts or specialists. 

At the moment we have had some progress in this direction but we still need more work toward a method for the specification of fault tolerant systems. The basic idea behind \cite{JonesHJ07} was to specify a system not in isolation but considering the environment in which it is going to run and deriving the final specification from a wider system where assumptions have been understood and formalized as layers of rely conditions. Here the difference between assumptions and requirements is crucial, especially when considering the proper fault tolerance aspects. We could briefly summarize this philosophy as follows:

\begin{itemize}
\item Not specifying the digital system in isolation
\item Deriving the specification starting from a wider system in which physical phenomena are measurable
\item Assumptions about the physical components can be recorded as layers of rely-conditions (starting with stronger assumptions and then weakening when faults are considered)
\end{itemize}

\begin{figure}[htp]
\centering
\includegraphics[scale=0.2]{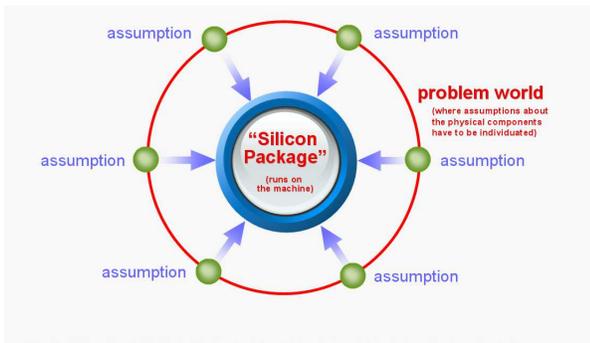}
\caption{Silicon Package, Problem World and Assumptions}\label{problem}
\end{figure}

Sometimes we have found useful, in the presentation of these concepts, to use figure \ref{problem}. This figure allows us to show how a computer system can be seen from a different angle, as not consisting of functions performing tasks in isolation but as relationships (interfaces/contracts) in a wider world including both the machine and the physical (measurable) reality. As we will see later, this philosophy has been inspired by Michael Jackson's approach to software requirements analysis typically called Problem Frames approach \cite{JacksonPF}. The Silicon Package is the software running on the hosting machine. It should be clear that the machine itself can neither acquire information on the reality around nor modify it. The machine can only operate trough sensors and actuators. To better understand this point, we like to use a similar metaphor about humans where it is easier to realize that our brain/mind system (our Silicon Package?) cannot acquire information about the world  but it can only do that through eyes, ears and so on (our sensors). In the same way it cannot modify the world if not through our arms, voice, etc (our actuators). So, as we start describing problems in the real world in terms of what we perceive and what we do (and not about our brain functioning) it makes sense to adopt a similar philosophy for computer systems consisting of sensors and actuators. Around the Silicon Package you can see a red circle representing the problem world and green small spheres representing the assumptions that need to be made regarding it. The arrows and their directions represent the fact that we want to derive the specification of the silicon package starting from the wider system. The way in which we record these assumptions is a topic for the following sections. 

\subsection*{The method, its Steps and its Views}

In \cite{mazzara:ewdc09} we analyzed the method introduced in \cite{JonesHJ07} according to the properties described in \cite{Descartes}. To do that, we recognized three macroscopic steps:

\begin{enumerate}
	\item Define boundaries of the systems
	\item Expose and record assumptions
	\item Derive the specification
\end{enumerate}

\begin{figure}[htp]
\centering
\includegraphics[scale=0.20]{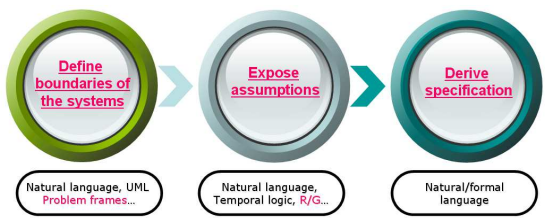}
\caption{Steps and Reference Tools}\label{phases}
\end{figure}

Our idea is not committing to a single language/notation - we want a formal method, not a formal language - so we will define a general high level approach following these guidelines and we will suggest \textit{reference tools} to cope with these steps. It is worth noting that these are only reference tools that are \textit{suggested} to the designers because of a wider experience regarding them from our side. A formal notation can be the final product of the method but it still needs to be not confused with the method itself. In figure \ref{phases}, these steps are presented and it is showed how different tools could fit the method at different stages. We call these notations the plug-ins since they can be plugged into the steps.

Figure \ref{phases} is a generic representation of the method where we want to emphasize the different steps which were not clearly defined in \cite{JonesHJ07}. The reader will understand that this is still a simplification of the process. We use the word "steps" instead of "phases" since we do not want to suggest a sort of linear process which is not always applicable, especially when coping with fault tolerance (as we will discuss later). We imagine, in the general case, many iterations between the different steps. The idea of the method is to ground the view of the silicon package in the external physical world. This is the problem world where assumptions about the physical components \textit{outside} the computer itself have to be recorded. Only after this can we derive the specification for the software that will run \textit{inside} the computer. A more precise formalization of the method and the features it has to exhibit is one of the main contributions of \cite{mazzara:ewdc09}. The reader is probably realizing that what we are obtaining here is a method exploiting two different perspectives during the three steps. 

\begin{itemize}
	\item a \textit{static view} defining the boundaries of the system and representing the relationships between phenomena and domains in it. Our reference tools here are Problem Diagrams \cite{JacksonPF}. 
	\item a \textit{dynamic view} representing the interactions between different processes in the system and able to record the assumptions. Our mathematical reference tools here are rely/guarantee conditions \cite{Jones83a,Jones83b} which regard the execution of concurrently executing (and interfering) processes.
\end{itemize}

Furthermore we need an approach to consider faulty behavior. This will be described later in the related section. The idea behind having two different views is that different people (or stakeholders) could possibly be interested only in single aspects of the specification and be able to understand only one of the possible projections. In this way you can approach the specification without a full understanding of every single aspect.

\subsection*{Static View}

Michael Jackson is well known  for having pioneered, in the seventies (with Jean-Dominique Warnier and Ken Orr) the technique for structuring programming basing on correspondences between data stream structure and program structure \cite{Jackson75}. Jackson's ideas acquired then the acronym JSP (Jackson Structured Programming). In his following contribution \cite{Jackson83} Jackson extended the scope to systems. Jackson System Development (JSD) already contained some of the ideas that made object-oriented program design famous. 

In this section we describe our reference tool for  representing the relationships between phenomena and domains of the system we want to specify using Problem Diagrams \cite{JacksonPF}. Context Diagrams and Problem Diagrams are the graphical notations introduced by Michael Jackson (in the time frame 1995/2001) in his Problem Frames (PF) approach to software requirements analysis. This approach consists of a set of concepts for gathering requirements and creating specifications of software systems. As previously explained, the new philosophy behind PF is that user requirements are here seen as being about relationships in the operational context and not functions that the software system must perform. It is someway a change of perspective with respect to other requirements analysis techniques. 

The entire PF software specification goal is modifying the world (the problem environment) through the creation of a dedicated machine which will be then put into operation in this world. The machine will then operate bringing the desired effects. The overall philosophy is that the problem is located in the world and the solution in the machine. The most important difference with respect to other requirements methodologies is the emphasis on describing the environment and not the machine or its interfaces. Let us consider, for example, the Use Case approach \cite{BittnerUC}. What is done here is specifying the interface, the focus is on the interaction user-machine. With PF we are pushing our attention beyond the machine interface, we are looking into the real world. The problem is there and it is worth starting there. The first two points of the ideas taken from \cite{JonesHJ07} (not specifying the digital system in isolation and deriving the specification starting from a wider system in which physical phenomena are measurable) can be indeed tracked back, with some further evolution, to \cite{JacksonPF}. In this work, we are using PF to develop a method for specification of systems, i.e. a description of the machine behavior. But, before doing that, we need to start understanding the problem. 

\subsection*{Context Diagrams}
The modeling activity of a system should start using this kind of diagram in the PF philosophy. By means of it we are able to identify the boundaries of the system, where a system is intended as the machine to be designed (software + hardware) and its domains with their connections (in terms of shared phenomena). It is part of what we call a static view of the system.

Context Diagrams contain an explicit and graphical representation of:
\begin{itemize}
	\item the machine to be built 
	\item the problem domains that are relevant to the problem
	\item the interface (where the Machine and the application domain interact)
\end{itemize}

\begin{figure}[htp]
\centering
\includegraphics[scale=0.25]{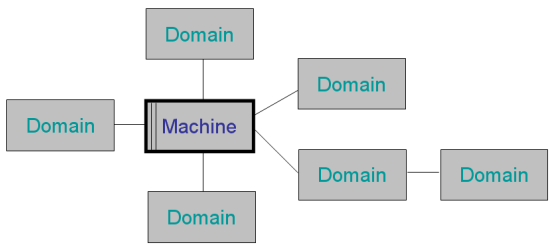}
\caption{Context Diagram}\label{CD}
\end{figure}

A domain here is considered to be a part of the world we are interested in (phenomena, people, events). A domain interface is where domains communicate. It does not represent data flow or messages but shared phenomena (existing in both domains). Figure \ref{CD} shows a simple scenario. The lines represent domain interfaces, i.e. where domains overlap and share phenomena.

\subsection*{Problem Diagrams}
  
The basic tool for describing a problem is a Problem Diagram which can be considered a refinement of a Context Diagrams. This should be the 2nd step of the modeling process. A problem diagram shows the requirements on the system, its domains, and their connections. It is still part of a static view of the system but better represents the assumptions about the system and its environment. They are basic tools to describe problems. To the information contained in context diagrams they add:

\begin{itemize}
	\item dotted oval for requirements 
	\item dotted lines for requirements references 
\end{itemize}
  
\begin{figure}[htp]
\centering
\includegraphics[scale=0.25]{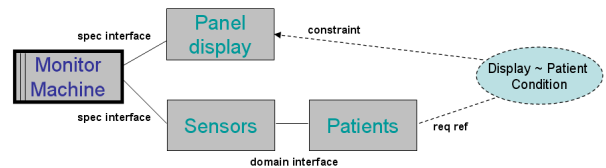}
\caption{Problem Diagram}\label{PD}
\end{figure}

Figure \ref{PD} shows a scenario where the Silicon Package is in charge of monitoring the patients conditions. We believe that the first step of the specification method (define boundaries of the systems) can be accomplished by means of this tools. Thus we use Problem Diagrams as a reference tool for our research but still, as said, not constraining it to a specific notation or language. 
 
\subsection*{Dynamic View}

Problem Diagrams taken from the PF approach are a notation that forces us to think about the problem in the physical world instead of focusing immediately on the solution. We believe that they represent an effective tool to define the precise boundaries of the specification we are working on. Summarizing they represent:

\begin{enumerate}
\item the machine
\item the problem domains
\item the domain interfaces
\item the requirements to bring about certain effects in the problem domains 
\item references in the requirements to phenomena in the problem domains
\end{enumerate}

Once the domains of the context we are working on, their phenomena and the relative overlap have been understood, it will be necessary to focus on the "border" between the Silicon Package and the real world. It is necessary to distinguish between assumptions and requirements and we need a tool to record assumptions. Our system will be composed of interacting parts and each of these parts will also interact with the world. The world itself has to be understood in term of assumptions about normal/abnormal behavior and a model of fault need to be considered. For all these reason we introduce the concept of \textit{dynamic view} which represents the interactions between processes in the system and between the system and the world. To record our assumption (as we will see layers of assumption for fault tolerance) we use a mathematical reference tool, i.e. rely/guarantee conditions \cite{Jones83a,Jones83b} which regard the execution of concurrently executing processes. R/G conditions are a powerful abstraction for reasoning about interference and they originated in the Hoare logic idea of preconditions and postcondition \cite{Hoare83}. The purpose is providing a set of logical rules for reasoning about the correctness of programs. We will explain the idea through examples, for more details please consider the literature. As the reader will realize in this section, rely conditions can be used to record assumptions in the overall context of the proposed method. However, as stated in \cite{CJ00}, when they show too much complication this might be a warning indicating a messy interface.

\subsection*{Preconditions and Postcondition}

To understand the power of the R/G reasoning it is necessary to realize how preconditions and postconditions can help in specifying a software program when interference does not play its role. What we have to describe (by means of logical formulas) when following this approach is:

\begin{enumerate}
	\item the input domain and the output range of the program
	\item the precondition, i.e. the predicate that we expect to be true at the beginning of the execution
	\item the postcondition, i.e. the predicate that will be true at the end of the execution provided that the precondition holds 
\end{enumerate}

Preconditions and postconditions represent a sort of contracts between parties: provided that you (the environment, the user, another system) can ensure the validity of a certain condition, the implementation will surely modify the state in such a way that another known condition holds. There is no probability here, it is just logic: if this holds that will hold. And the input-output relation is regulated by a predicate that any implementation has to satisfy.

We show the example of a very simple program, the specification of which in the natural language may be: \textit{``Find the smallest element in a set of natural numbers''}.

This very simple natural language sentence tells us that the smallest element has to be found in \textit{a set of natural numbers}. So the output of our program has necessarily to be a natural number. The input domain and the output range of the program are then easy to describe:

$$I/O:  \mathcal{P}(\textbf{N}) \rightarrow \textbf{N}$$ 

Now, you expect your input to be a set of natural numbers, but to be able to compute the min such a set has to be non empty since the min is not defined for empty sets. So the preconditions that has to hold will be:

$$P(S):  S \neq \emptyset$$

Provided that the input is a set of natural numbers \textit{and} it is not empty, the implementation will be able to compute the min element which is the one satisfying the following:

$$Q(S,r): r \in S  \wedge (\forall e \in S)(r \leq e) $$

Given this set of rules, the input-output relation is given by the following predicate that needs to be satisfied by any implementation \textit{f}:

$$\forall S \in \mathcal{P}(\textbf{N}) (P(S) \Rightarrow f(S) \in  \textbf{N} \wedge Q(S, f(S)))$$ 

\subsection*{Interference}

\begin{figure}[htp]
\centering
\includegraphics[scale=0.2]{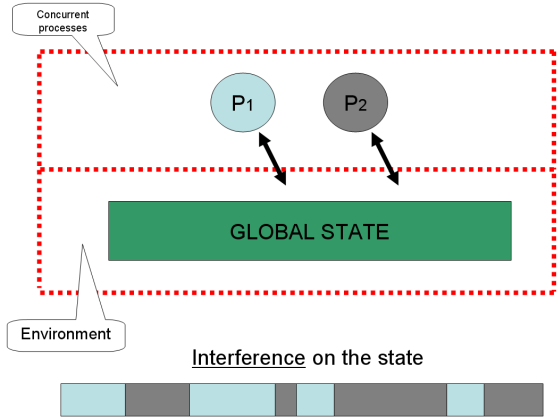}
\caption{Interference trough global state}\label{int}
\end{figure}

The example just showed summarizes the power (and the limitations) of this kind of abstractions. To better understand the limitations consider figure \ref{int} where interference and global state are depicted. The two processes alternate their execution and access the state. The global state can consist of shared variables or can be a queue of messages if message passing is the paradigm adopted. This figure shows exactly the situations described in \cite{Jones83a}, quoting precisely that work:

\begin{quote}
As soon as the possibility of other programs (processes) running in parallel is admitted, there is a danger of "interference." Of more interest are the places where it is required to permit parallel processes to cooperate by changing and referencing the same variables. It is then necessary to show that the interference assumptions of the parallel processes coexist.
\end{quote}

Another quote from \cite{ColletteJ00} says:

\begin{quote}
The essence of concurrency is interference: shared-variable programs must be designed so as to tolerate state changes; communication-based concurrency shifts the interference to that from messages. One possible way of specifying interference is to use rely/guarantee-conditions.
\end{quote}

In case we consider interfering processes, we need to accept that the environment can alter the global state. However,the idea behind R/G is that we impose these changes to be constrained. Any state change made by the environment (other concurrent processes with respect to the one we are considering) can be assumed to satisfy a condition called R (rely) and the process under analysis can change its state only in such a way that observations by other processes will consist of pairs of states satisfying a condition G (guarantee). Thus, the process \textit{relying} on the fact that a given condition holds can \textit{guarantee} another specific condition. An example is now presented.

\subsection*{Greatest Common Divisor}

Consider the two following simple pieces of code, the cooperation of which calculates the Greatest Common Divisor:

\small
$$
\begin{array}{lll}
\texttt{P1:} 					               & \; \; \; \; \; \; \; \; \; \; \; \; \; \; \; \; \; \; \; \; \; \; \; \; & \texttt{P2:} \\
\texttt{while(a<>b)} \texttt{\{}     & \; \; \; \; \; \; \; \; \; \; \; \; \; \; \; \; \; \; \; \; \; \; \; \; & \texttt{while(a<>b)\{} \\
\; \; \; \;  \texttt{if(a > b)}      & \; \; \; \; \; \; \; \; \; \; \; \; \; \; \; \; \; \; \; \; \; \; \; \; & \; \; \; \; \texttt{if(b > a)} \\
\; \; \; \; \; \; \texttt{a := a-b;} & \; \; \; \; \; \; \; \; \; \; \; \; \; \; \; \; \; \; \; \; \; \; \; \; & \; \; \; \; \; \;  \texttt{ b := b-a;}  \\
\texttt{\}} 					               & \; \; \; \; \; \; \; \; \; \; \; \; \; \; \; \; \; \; \; \; \; \; \; \; & \texttt{\}}				 
\end{array}
$$

\normalsize
\noindent

P1 is in charge of decrementing \textit{a} and P2 of decrementing \textit{b}. When $a=b$ will evaluate to true it means that one is the Greatest Common Divisor for \textit{a} and \textit{b}. The specification of the interactions is as follows:\\

\noindent
\small
$R_1:(a=\overline{a}) \wedge (a \geq b \Rightarrow b= \overline{b}) \wedge (GCD(a,b) = GCD(\overline{a},\overline{b}) )$\\
$G_1:(b=\overline{b}) \wedge (a \leq b \Rightarrow a=\overline{a}) \wedge (GCD(a,b) = GCD(\overline{a},\overline{b}) ) $\\
$R_2 = G_1$\\
$G_2 = R_1$\\

\normalsize
\noindent
Here the values $\overline{a}$ and $\overline{b}$ are used instead of \textit{a} and \textit{b} when we want to distinguish between the values before the execution and the values after. P1 relies on the fact that P2 is not changing the value of \textit{a} and $a \geq b$ means no decrements for  \textit{b} have been performed. Furthermore the CGD did not change. Specular situation is for the guarantee condition. Obviously, what is a guarantee for P1 becomes a rely for P2 and vice versa.

\subsection*{Need for Extension (of Jackson's Diagrams)?} 

The objective of a PF analysis is the decomposition of a problem into a set of subproblems, where each of these matches a problem frame. A problem frame is a problem pattern, i.e the description of a simple and generic problem for which the solution is already known. There are four main patterns plus some variations:

\begin{itemize}
	\item required behavior (the behavior of a part of the physical world has to be controlled)
	\item commanded behavior (the behavior of a part of the physical world has to be controlled in accordance with commands issued by an operator)
	\item information display (a part of the physical world states and behavior is continuously needed)
	\item simple workpieces (a tool is needed for a user to create/edit a class of text or graphic objects so that they can be copied, printed...)
\end{itemize}

Our perception is that, when describing the behavior of interfering processes - especially when faults are considered as a special case of interference (see next section) - the diagrams and the patterns provided are not powerful enough. We need further refinement steps filling the gap between the static and the dynamic view to complete the specification process. Now we briefly describe these ideas that needs further work and can be considered an open issue.

\subsection*{Interface Diagram}
In a 3rd step of the modeling process, we want to represent an external, static view of the system. We need a further refinement of the Problem Diagram able to identify the operations of the system and its domains, and the input/output data of these operations (with their types). The relationship of these with the requirements identified in the Problem Diagram has to be represented at this stage.

\subsection*{Process Diagram}
In a 4th step of the modeling process, the whole system is represented as a sequential process and each of its domains as a sequential process. Concurrency within the system or within its domains is modeled by representing these as two or more subcomponents plus their rely and guarantee conditions. This is an external, dynamic view of the system and its domains.

\section{An Angle to See Faults}

Testing can never guarantee that software is correct. Nevertheless, for specific software features - especially the ones involving human actions and interactions - rigorous testing still remains the best choice to build the desired software. We know very little about human behavior, there are few works trying to categorize, for example, human errors in such a way that we can design system that can prevent bad consequences \cite{ReasonHR} but this goes far beyond the scope of this work. Here we want to focus on the goal of deploying highly reliable software in terms of aspects that can be quantified (measured), for example the functional input/output relation (or input/output plus interference, as we have seen). In this case formal methods and languages provide some support. The previous sections discussed how to derive a specification of a system looking at the physical world in which it is going to run. No mention has been made of fault tolerance and abnormal situations which deviate from the basic specification. The reader will soon realize that the method we have defined does not directly deal with these issues but it does not prevent fault tolerance from playing a role. The three steps simply represent what you have to follow to specify a system and they do not depend on what you are actually specifying. This allows us to introduce more considerations and to apply the idea to a wider class of systems. Usually, in the formal specification of sequential programs, widening the precondition leads to make a system more robust. The same can be done weakening rely conditions. For example, if eliminating a precondition the system can still satisfy the requirements this means we are in presence of a more robust system. In this paper we will follow this approach presenting the notion of Layered Fault Tolerant Specification (LFTS) and examining the idea of fault as interference \cite{ColletteJ00}, i.e. a different angle to perceive system faults. Quoting \cite{ColletteJ00}:

\begin{quote}
The essence of this section is to argue that faults can be viewed as interference in the same way that concurrent processes bring about changes beyond the control of the process whose specification and design are being considered.
\end{quote}

The idea of Layered Fault Tolerant Specification (LFTS) is now presented in combination with the approach quoted above making use of rely/guarantee reasoning. The principle is layering the specification, for the sake of clarity, in (at least) two different levels, the first one for the \textit{normal behavior} and the others (if more than one) for the \textit{abnormal}. This approach originated from the notion of idealized fault tolerant component \cite{AndersonFT} but the combination of LFTS and rely guarantee reasoning can be considered one of the main contributions of this work. 

\subsection*{Fault Model}

First, when specifying concurrent (interfering) processes, we need to define which kind of abnormal situations we are considering. We basically need to define a Fault Model, i.e.what can go wrong and what cannot. Our specification will then take into account that the software will run in an environment when specific things can behave in an "abnormal" way.
There are three main abnormal situations in which we can incur, they can be considered in both the shared variables and message passing paradigm: 

\begin{itemize}
	\item Deleting state update: ``lost messages''
	\item Duplicating state update: ``duplicated messages''
	\item Additional state update (malicious): ``fake messages created''
\end{itemize}

The first one means that a message (or the update of a shared variable) has been lost, i.e. its effect will not be taken into account as if it never happened. The second one regards a situation in which a message has been intentionally sent once (or a variable update has been done once) but the actual result is that it has been sent (or performed) twice because of a faulty interference. The last case is the malicious one, i.e. it has to be done intentionally (by a human, it cannot happen only because of hardware, middleware or software malfunctioning). In this case a fake message (or update) is created from scratch containing unwanted information.

Our model of fault is represented by a so called \textit{Error Injector} (EI). The way in which we use the word here is different with respect to other literature where Fault Injector or similar are discussed. Here we only mean a model of the erroneous behavior of the environment. This behavior will be limited depending on the number of abnormal cases we intend to consider and the EI will always play its role respecting the RG rules we will provide. In the example we will show in the following we are only considering the first of the three cases, i.e the Fault Injector is only operating through lost messages.

A contribution of this work is the organization of the specification in terms of layers of Rely/Guarantee conditions. In order to do this we introduce the idea of EI as a model of the environment and we need to describe how the EI will behave and how we can limit it. Here a process will rely on a specific faulty behavior and, given that, will guarantee the ability to handle these situations. More in detail:

\begin{itemize}
	\item Rely: the Error Injector (environment) interferes with the process (changing the global state) respecting his G (superset of the program's R) --- for example, only ``lost messages'' can be handled (next example)
	\item Guarantee: The process provided this kind of (restricted) interference is able to handle exceptional/abnormal (low frequency) situations
\end{itemize}

All the possibilities of faults in the system are described in these terms and the specification is organized according to the LFTS principle we are going to describe.

\subsection*{LFTS: how to organize a clear specification}

The main motto for LFTS is: "Do not put all in the normal mode". From the expressiveness point of view, a monolithic specification can include all the aspects, faulty and non faulty of a system in the same way as it is not necessary to organize a program in functions, procedures or classes. The matter here is pragmatics, we believe that following the LFTS principles a specification can be more understandable for all the stakeholders involved.

The specification has to be separated in (at least) two layers, one for the \textit{Normal Mode} and one (or more) for the \textit{Abnormal Mode}. More specifically:

\begin{itemize}
	\item Normal mode: an operation usually runs in normal mode respecting his ``interface'' with the world determined by P/Q
	\item Fault interference: in ``low'' frequency cases the abnormal mode is ``activated'' (exception handler, forward recovery)
\end{itemize}

\begin{figure}[htp]
\centering
\includegraphics[scale=0.2]{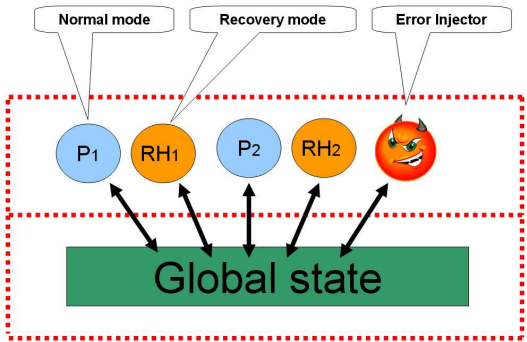}
\caption{Error Injector}\label{EI}
\end{figure}

Figure \ref{EI} shows the organization of a process (dashed rectangles) in a main part and a \textit{recovery handler} part where both interact through the global state with other processes and the Error Injector (represented by a devil here).

It is worth noting the limitations of this way of operating. Self error detection and self recovery cannot be addressed by this model since EI is a representation of the environment external to the process itself. So faulty behavior due to internal malfunctioning is not what we want to represent here.

\subsection*{Example of Specification of Interference}

For a better understanding of how we can exploit this idea of treating faults as extraordinary interference with a low frequency, we introduce a very simple example. First we consider an even simpler example without interference, then we introduce interference to investigate the differences and how we cope with them. 

\subsubsection*{Increments without Interference}
Let us consider the following piece of code:

\small
\begin{verbatim}
C(n):
n' := n;
while (n'>0){
      n' := n'-1; 	 
      count ++
}
return count;
\end{verbatim}
\normalsize

C is a very simple program which decrements its input while reaching zero. While decrementing the input it increments a counter with the effect that, at the end, the counter will obviously reach the original value of the input. The specification of C in terms of Pre and Post conditions is given as follows:

$$I/O:\textbf{N} \rightarrow \textbf{N}$$ 

The input (\textit{n}) and the output (\textit{count}) are natural numbers. The precondition that has to hold is:

$$P(count): count = 0$$

since we expect the counter to be zero at the beginning. Provided that the input is a natural number \textit{and} the counter is zero, the execution will satisfy the following:

$$Q(n, count): count = n \wedge \overline{n} = 0 $$

Without any interference, the specification of C only requires that the input-output relation satisfy the predicate:

$$\forall a \in \textbf{N} (P(a) \Rightarrow C(a) \in  \textbf{N} \wedge Q(a, C(a)))$$ 

\subsubsection*{Increments with Faulty Interference}
Let us consider the same piece of code:

\small
\begin{verbatim}
C(n):
n':= n;
while (n'>0){
      n' := n'-1; 	 
      count ++
}
return count;
\end{verbatim}
\normalsize

but running in an environment where the following EI is also running:

\small
\begin{verbatim}
EI(n'):
if (n'>0){
    n' := n'+1;
}
\end{verbatim}
\normalsize

The role of this EI here is to model the deletion of state updates as in the first of the three cases discussed above. The specification of C as expressed so far is too simple to be able to manage this kind of situations. Even if we are not handling malicious updates, the basic formulation we provided so far needs to be properly incremented because without any changes the desired implication cannot be satisfied:

$$\forall a \in \textbf{N} (P(a) \not \Rightarrow C(a) \in  \textbf{N} \wedge Q(a, C(a)))$$ 

What we have to do is restructure the implementation and to pass from pre and post conditions to rely/guarantee in the specification. Let us consider the following modification:

\small
\begin{verbatim}
C(n):
n':= n;
while (n'>0){
   if n'+ count = n then {       			
	    n' := n'-1;
	    count ++
	 } else { 
	    n' := n-count-1 
	}
}
return count;
\end{verbatim}
\normalsize

As the reader will understand what we have done is simply add a recovery handler and a recovery mode based on the evaluation of the condition $n'+ count = n$ which is able to flag the presence of an unwanted interference (a deletion of an increment). The recovery block is able to cope with abnormal situations provided that faults are restricted in behavior (and that it is known in advance). Thus, provided that a restricted interference happens the program is still able to satisfy the post condition (and the specification). The normal mode here is the simple code:
\begin{verbatim}
n' := n'-1; 	    		
count ++
\end{verbatim}
while the recovery handler is 
\begin{verbatim}
n' := n-count-1 
\end{verbatim}
and, as represented in figure \ref{EI}, C is running in an environment which is shared with EI. The specification we want in this case is different from the previous one and it is expressed, in terms of R/G conditions, as follows:
\\\\
\small
$R_C:(\overline{n} = n) \wedge ( \overline{count} = count) \wedge (\overline{n'} > n')$\\
$G_C: n' = n-count-1 $\\
$R_I = true $\\
$G_I = n' > \overline{n'} $
\\\\
\normalsize
It is worth noting that there is no rely condition (to be precise there is one always true) for the Error Injector, indeed it would not be reasonable to expect that the processes we are specifying would behave in a way so as to satisfy the needs of a fault model. Instead, EI is guaranteeing that it will only increment $n'$ - it is the case of having only state update deletion (an increment deletes a decrement) as pointed out previously. Decided the EI behavior limitation (and thus decided the fault model) we can design our specification. From the EI specification C can rely on the fact that \textit{n} and \textit{count} will be never modified while \textit{n'} will be only modified in a specific way (incremented). Now, with the addition of a layer in the program and in the specification we are still able to guarantee an (extended) desired behavior by means of the $G_C$ condition which says that \textit{n'} will always be consistent with the value of \textit{count} preserving the invariant $n' = n-count-1 $, i.e. the summation of  \textit{n'} and  \textit{count} will always be equal to $n-1$. This will ensure that the postcondition  $count = n \wedge \overline{n} = 0$ will hold at the end like in the case without interference. This simple example shows how the LFTS principles can provide a clear specification (with respect to a monolithic one) ensuring, at the same time, that a desired postcondition holds.
    
\section{The Automotive Example}

The progress toward this way of layering specifications has been made by experimenting few case studies. For example, the one presented in \cite{mazzara:ewdc09} showed the power of the LFTS principle when applied to train systems. Instead we now consider a simplified automotive case study. The Cruise Control is a system able to automatically control the rate of motion of a motor vehicle. The driver sets the speed and the system will take over the throttle of the car to maintain the same speed. One of the requirements of the cruise control is to be switched off when an error in the engine speed sensor is detected. This has to be taken into account in the specification. We use the CrCt to show how the idea of LFTS can be applied in (semi)realistic systems (simplifications of real system for the sake of experimenting with new ideas but still not mere toy examples). Let us consider the following ideal piece of CrCt code:

\small
\begin{verbatim}
while (target <> current){
   delta := smooth(target, current); 	
   result := set_eng(delta);
}
\end{verbatim}
\normalsize

%\begin{figure}[htp]
%\centering
%\includegraphics[scale=0.2]{CrCt.eps}
%\caption{Simplified State Machine (no tip up/down)}\label{CrCt}
%\end{figure}

The car speed is acquired in \texttt{smooth(target, current)} and then a delta is calculated for the car to have a smooth acceleration (smoothness has to be determined by experience). The specification of this code in term of P,Q,R,G is the following (it is expressed in natural language since we are not giving a mathematical model of the car here):\\
\begin{itemize}
	\item P: target has to be in a given range
	\item Q: delta is zero and the driver has been comfortable with the acceleration
	\item R: the engine is adjusted (smoothly) according to delta 
	\item G: the absolute value of delta is decreasing\\
\end{itemize}

The requirement mentioned above is not taken into account in this ideal piece of code, so in case the speed acquisition goes wrong the guarantee will not hold and the absolute value of delta will not be decreased. Indeed, following the LFTS principle we should organize it in two layers: a normal mode and an abnormal one (speed acquisition goes wrong): 

\small
\begin{verbatim}
while (target <> current){
   delta := smooth(target, current); 	
   result := set_eng(delta);
   if result <> OK then 
      switch_off
}
\end{verbatim}
\normalsize

This means adding a weaker layer of conditions for the ``abnormal case'' being still able to guarantee ``something''. If speed acquisition goes wrong we do not want to force the engine following the delta since it would imply asking for more power when, for example, the car speed is actually decreasing  (maybe an accident is happening or it is just out of fuel). Switching the engine off we avoid an expensive engine damage. %Figure \ref{layers} summarizes the layering and the entire structure of this simple example. 

%\begin{figure}[htp]
%\centering
%\includegraphics[scale=0.2]{layers.eps}
%\caption{Layered CrCt Specification}\label{layers}
%\end{figure}

%In this paper Jackson's diagrams extensions like Interference Diagrams and Process Diagrams are not fully developed, anyway an ideal Process Diagram for the Cruise Control is depicted in Figure \ref{CrCtviews}. For the diagram to be complete we should be able to describe the rely/guarantee conditions of each subcomponents represented, anyway this is not a goal of the paper and it would face confidentiality issues related to the project. In its current sanitized state this diagram is able to show how a Problem Diagram is linked to the dynamic set of existing processes, how they belongs to different domains and how the requirements are related to this.
%
%\begin{figure}[htp]
%\centering
%\includegraphics[scale=0.2]{CrCtviews.eps}
%\caption{Static and Dynamic views combined}\label{CrCtviews}
%\end{figure}
   
\section{Conclusive Remarks}

In this work we provided a different view for interpreting problems and faults and we worked toward an improvement of the ideas presented in \cite{JonesHJ07}. Our goal was to start an investigation leading to a method for the formal specification of systems that do not run in isolation but in the real, physical world. To accomplish the goal we passed trough a non trivial number of steps including the discussion in \cite{mazzara:ewdc09} of the concept of method itself (computer science has a proliferation of languages but very few methods). Then we presented how we intend to proceed to represent the static and the dynamic view of the problem. A section is dedicated to faults and the following to a case study. 

Of course this work is not exhaustive and many aspects need more investigation. Especially the possibility of having Jackson's diagrams extensions working as a bridge between the static and the dynamic view in the way we described them. Although a small example of static and dynamic views is presented in this paper and a way to combine them idealized, more work is needed in combining them in a coherent and readable notation. Jackson' diagrams extensions are only one of the possible solutions anyway. Indeed another point we have just sketched here but that needs more work is about the the plug-ins and how to permit the practical use of different tools/notation. More investigation regarding the case studies is also needed.

\small

\section*{Acknowledgments} This work has made been possible by the useful conversations with Cliff Jones, Michael Jackson, Ian Hayes, Ani Bhattacharyya, Alexander Romanovsky, John Fitzgerald, Jeremy Bryans, Fernando Dotti, Alexei Iliasov, Ilya Lopatkin, Rainer Gmehlich and Felix Loesch and it has been funded by the EU FP7 DEPLOY Project \cite{DEPLOY}.

\bibliographystyle{abbrv}
\bibliography{refs}

\end{document}